\documentclass[trackchanges,twocolumn]{aastex701}

\usepackage{multirow}

\begin{document}

\title{Sub-Neptunes Show a Stronger Correlation with Cold Jupiters than Super-Earths Especially in Metal-rich Systems}

\correspondingauthor{Di-Chang Chen}
\email{chendch28@mail.sysu.edu.cn}

\author[orcid=0000-0003-0707-3213,sname='Chen']{Di-Chang Chen}
\affiliation{School of Physics and Astronomy, Sun Yat-sen University, Zhuhai 519082, China}
\affiliation{CSST Science Center for the Guangdong-Hong Kong-Macau Great Bay Area, Sun Yat-sen University, Zhuhai 519082, China}
\email{chendch28@mail.sysu.edu.cn}  


\author[orcid=0000-0002-8958-0683]{Fei Dai}
\affiliation{Institute for Astronomy, University of Hawai'i, Honolulu, HI 96822, USA}
\email{fdai@hawaii.edu}

\author{Bo Ma}
\affiliation{School of Physics and Astronomy, Sun Yat-sen University, Zhuhai 519082, China}
\affiliation{CSST Science Center for the Guangdong-Hong Kong-Macau Great Bay Area, Sun Yat-sen University, Zhuhai 519082, China}
\email{mabo8@mail.sysu.edu.cn}

\author{Shang-Fei Liu}
\affiliation{School of Physics and Astronomy, Sun Yat-sen University, Zhuhai 519082, China}
\affiliation{CSST Science Center for the Guangdong-Hong Kong-Macau Great Bay Area, Sun Yat-sen University, Zhuhai 519082, China}
\email{liushangfei@mail.sysu.edu.cn}

\author{Cong Yu}
\affiliation{School of Physics and Astronomy, Sun Yat-sen University, Zhuhai 519082, China}
\affiliation{CSST Science Center for the Guangdong-Hong Kong-Macau Great Bay Area, Sun Yat-sen University, Zhuhai 519082, China}
\email{yucong@mail.sysu.edu.cn}
 
\begin{abstract}
Correlations between the inner small planets and cold giants encodes the formation and evolution of planetary systems.
It remains unclear if the correlation differs on the two sides of the radius valley.
In this work, we compute the conditional frequency of cold Jupiters in systems with only inner sub-Neptunes $P(\rm CJ|SN)$ and those with only inner super-Earths $P(\rm CJ|SE)$.
We find that, around transiting sample around metal-rich stars, $P(\rm CJ|SN, [Fe/H]>0)$ and $P(\rm CJ|SE, [Fe/H]>0)$ are $42.6^{+10.6}_{-9.9}\%$ and $14.5^{+12.7}_{-6.9}\%$.
Comparing with the field giant frequency ($14.3^{+2.0}_{-1.8}\%$), we show that inner sub-Neptunes and cold Jupiters exhibit a significant positive correlation for metal-rich systems with a confidence level of 99.95\%, whereas this correlation is absent for systems with super-Earths.
{We also consider a homogeneous Kepler-Keck subsample and derive similar results, with $P(\rm CJ|SN, [Fe/H]>0)$ of $45.8^{+18.6}_{-16.3}\%$ and $P(\rm CJ|SE, [Fe/H]>0)$ of $13.3^{+17.0}_{-6.8}\%$.}
Radial velocity sample shows consistent results, with metal-rich systems hosting massive inner planets exhibiting a strong positive correlation (confidence level of 99.11\%) with outer cold Jupiters ($P(\rm CJ|M_{p}>10M_\oplus, [Fe/H]>0) = 34.6^{+11.0}_{-9.1}\%$).
{These results can be naturally understood since metal-rich disks are expected to more efficiently produce both outer cold Jupiters and inner planets with larger radii and masses.
Our findings highlight the critical role of stellar metallicity in shaping planetary architectures, particularly for large/massive planets.}

\end{abstract}



\section{Introduction}
\label{sec.intro}
Over the past two decades, exoplanet surveys have revealed that planets with sizes between those of Earth and Neptune are abundant in the Galaxy. Statistical studies based on radial velocity and transit observations suggest that $\sim 30\%-50\%$ Sun-like stars host at least one such planets in inner orbits \citep[e.g.,][]{2011arXiv1109.2497M,2018AJ....156...24M,2018ApJ...860..101Z}.
Understanding how they form and interact with other planetary systems, particularly outer gas giants, is therefore essential for developing a comprehensive theory of planet formation and evolution.

The correlation between inner small planets and outer cold Jupiters has been considered a useful test.
Using carefully constructed samples of radial velocity and transiting systems, \cite{2018AJ....156...92Z} reported that cold Jupiters occur in approximately one-third of systems with inner small planets, which is about three times higher than that around field stars. This excess of cold Jupiters in systems hosting inner small planets was later confirmed by several independent studies \citep[e.g.,][]{2019AJ....157...52B,2022ApJS..262....1R}. 
However, more recent analyses have suggested a weaker or even negative correlation between the two populations \citep[e.g.,][]{2023A&A...677A..33B}. 
A possible resolution to this discrepancy was proposed by \cite{2024RAA....24d5013Z}, who argued that stellar metallicity is a key parameter governing the observed correlation. 
Indeed, by analyzing the largest sample to date of systems hosting both inner small planets and outer gas giants, \cite{2024ApJ...968L..25B} demonstrated that the correlation is pronounced among metal-rich stars ($\rm [Fe/H]>0$) but statistically insignificant for metal-poor systems, lending strong support to this interpretation.

Statistical studies have revealed another key feature of close-in exoplanets, radius valley, a deficit of planets around $\sim 1.9 R_\oplus$ that separates compact, rocky super-Earths from lower-density sub-Neptunes \citep{2013ApJ...775..105O,2017AJ....154..109F}.
Several theoretical models have been proposed to explain its origin.
Some studies suggested that sub-Neptunes are born with a H/He dominated gas envelope and the radius
valley is a result of atmospheric loss driven by photoevaporation \citep{2013ApJ...775..105O,2017ApJ...847...29O,2018ApJ...853..163J} or core-powered mass loss \citep{2016ApJ...825...29G,2018MNRAS.476..759G}.
Alternatively, other studies have suggested that sub-Neptunes are intrinsically volatile-rich, and the radius valley arises naturally from planet formation and migration processes \citep{2019PNAS..116.9723Z,2020A&A...643L...1V,2021ApJ...908...32L,2024NatAs...8..463B}.
Plenty of studies have investigated the dependence of the radius valley on planetary and stellar properties \citep[e.g., orbital period, stellar mass, metallicity, age;][]{2018MNRAS.479.4786V,2018MNRAS.480.2206O,2020AJ....160..108B,2022AJ....163..249C}.
However, current observational constraints remain insufficient to conclusively distinguish among these competing theoretical scenarios.

In this work, we thus divide the inner planets into two categories (i.e., sub-Neptune, and super-Earth) and investigate their correlations with cold Jupiters based on the planetary sample in \cite{2024ApJ...968L..25B}, aiming to help constrain the formation and migration histories.
In section \ref{sec.sample}, we describe how we categorize the planetary systems.
In section \ref{sec.results}, we calculated the conditional frequency of cold Jupiters in different type of systems around metal-rich and metal-poor systems.
Finally, section \ref{sec.dis_con} summarizes our
findings and discusses their implications.

\section{Sample}
\label{sec.sample}

\cite{2024ApJ...968L..25B} collects a sample of 184 systems with publicly available radial velocity data and detected outer gas giants.
Since most small planets discovered by the radial velocity (RV) method lack radius measurements, and the subsequent classification relies on planetary radii, we therefore focus on systems with the 109 transiting small planets (i.e., $1 R_\oplus \le R_{\rm P} \le 4 R_\oplus$). 
Among these systems, 16 systems host one or more cold Jupiters, with masses of $0.5-20 M_{\rm J}$ and semi-major axes of $1-10$ AU.

Previous studies have found that the radius valley is located at $R_{\rm Valley}^0 \sim 1.9 \pm 0.2 R_\oplus$ \citep{2017ApJ...847...29O,2017AJ....154..109F} and is dependent on stellar mass $M_*$ \citep{2020AJ....160..108B} and planetary orbital period $P$ \citep{2018MNRAS.479.4786V}, which can be characterized as:
\begin{equation}
    R^{\rm valley} = R_{\rm valley}^0
    \left(\frac{M_*}{M_\odot}\right)^h
    \left(\frac{P}{\rm 10 days}\right)^g,
\end{equation}
where $h$ and $g$ are are the corresponding slopes, which are adopted as $0.21^{+0.06}_{-0.07}$ and $-0.09^{+0.02}_{-0.03}$ \citep{2023MNRAS.519.4056H,2024MNRAS.531.3698H}, respectively.

We then classify planets into two categories based on whether their radii lie above or below the planetary radius valley: sub-Neptunes and super-Earths.
Based on the planet categories of the small planets in each planetary system, we further divide the systems into three types:
\begin{enumerate}
    \item Sub-Neptune systems, i.e., systems with only sub-Neptunes: 52 systems, 10 of which hosting cold Jupiters.

    \item Super-Earth systems, i.e., systems with only super-Earths: 28 systems, 3 of which hosting cold Jupiters.
    \item Mixed systems, i.e., systems with both Sub-Neptunes and super-Earths.
    29 systems, 3 of which hosting cold Jupiters.
\end{enumerate}

\begin{figure*}[t]
\centering
\includegraphics[width=\linewidth]{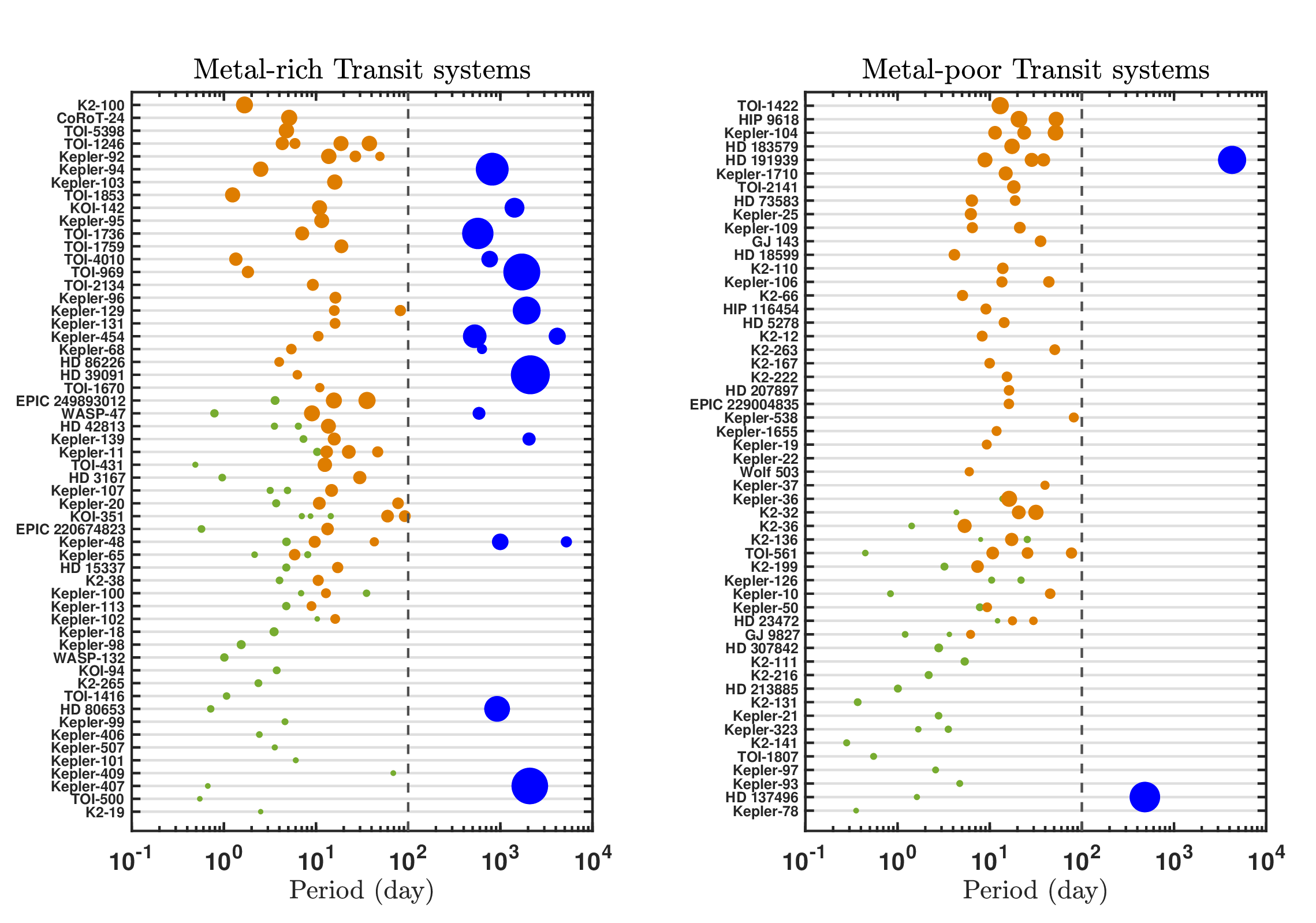}
\caption{Orbital architectures of the 56 metal-rich (Left) and 53 metal-poor (Right) transiting planetary systems. Super-Earths, sub-Neptunes and cold Jupiters are plotted in green, orange and blue, respectively. 
The three types (i.e., sub-Neptune, mixed and super-Earth) of systems are arranged from top to bottom. 
Within each type, the systems are sorted by the radius of the largest planet in the system.
The symbol sizes of Super-Earths and sub-Neptunes are proportional to the planetary cross-sectional areas, while those of cold Jupiters represent the planetary masses.
\label{figPRtransit}}
\end{figure*}

Figure \ref{figPRtransit} displays the period-radius diagram coloured by the planetary types for the three types of systems.

\section{Results}
\label{sec.results}

\subsection{Calculating the conditional frequency of Cold Jupiters}
\label{sec.results.method}
To assess the correlation between inner planets and cold Jupiters, we first calculate the conditional frequency of gas giants for the three system types: sub-Neptune, mixed, and super-Earth systems.
Referring to previous studies \citep{2024ApJ...968L..25B,2024RAA....24d5013Z}, we model the detections using binomial statistics and adopt a flat prior on the conditional frequency.
Given $n_{\rm det}$ detections and the total number of systems $n_{\rm tot}$, the posterior distribution of the conditional rate is given by:
\begin{equation}
    f (P;a,b) = \frac{1}{B(a,b)} P^{a-1} (1-P)^{b-1},
\end{equation}
where $B(a,b)$ is the beta function $a=n_{\rm det}+1$ and $b=n_{\rm eff}-n_{\rm det}+1$. 
$n_{\rm eff}$ is the effective total number after correcting for detection completeness to gas giants.
For the detection efficiency analysis, we collected the RV measurements and their uncertainties from the original literature. 
As shown in Figure \ref{figRVDE_SNSEMixtransiting} in Appendix A, the three types of systems exhibit similar distributions in {stellar mass, observational baselines, number of RV data points $N_{\rm obs}$ and RV uncertainties divided by $\sqrt{N_{\rm obs}}$ (with KS $p-$values $>0.3$) comparing to the whole transiting samples, and therefore have their detection completeness does not differ significantly}.
{It is worth noting that a non-significant K–S test result does not imply identical underlying populations, and could also be due to limited statistical power given the relatively small sample sizes.}
Thus we used the completeness maps in Figure A1 of \citep{2024ApJ...968L..25B} to determine the average completeness for the set of systems considered.
The frequency is reported as the maximum posterior probability estimate, and the uncertainties are defined as the 68\% confidence interval centered on this value. 

We then split the sample into metal-rich ($\rm [Fe/H]>0$) and metal-poor ($\rm [Fe/H] \le 0$) subsamples and compare the conditional and filed frequency of cold Jupiters within same metallicity range.

\subsection{Excess of Cold Jupiters in sub-Neptune systems around Metal-rich stars}
\label{sec.result.FCJ_rich}

\begin{figure}[t]
\centering
\includegraphics[width=\linewidth]{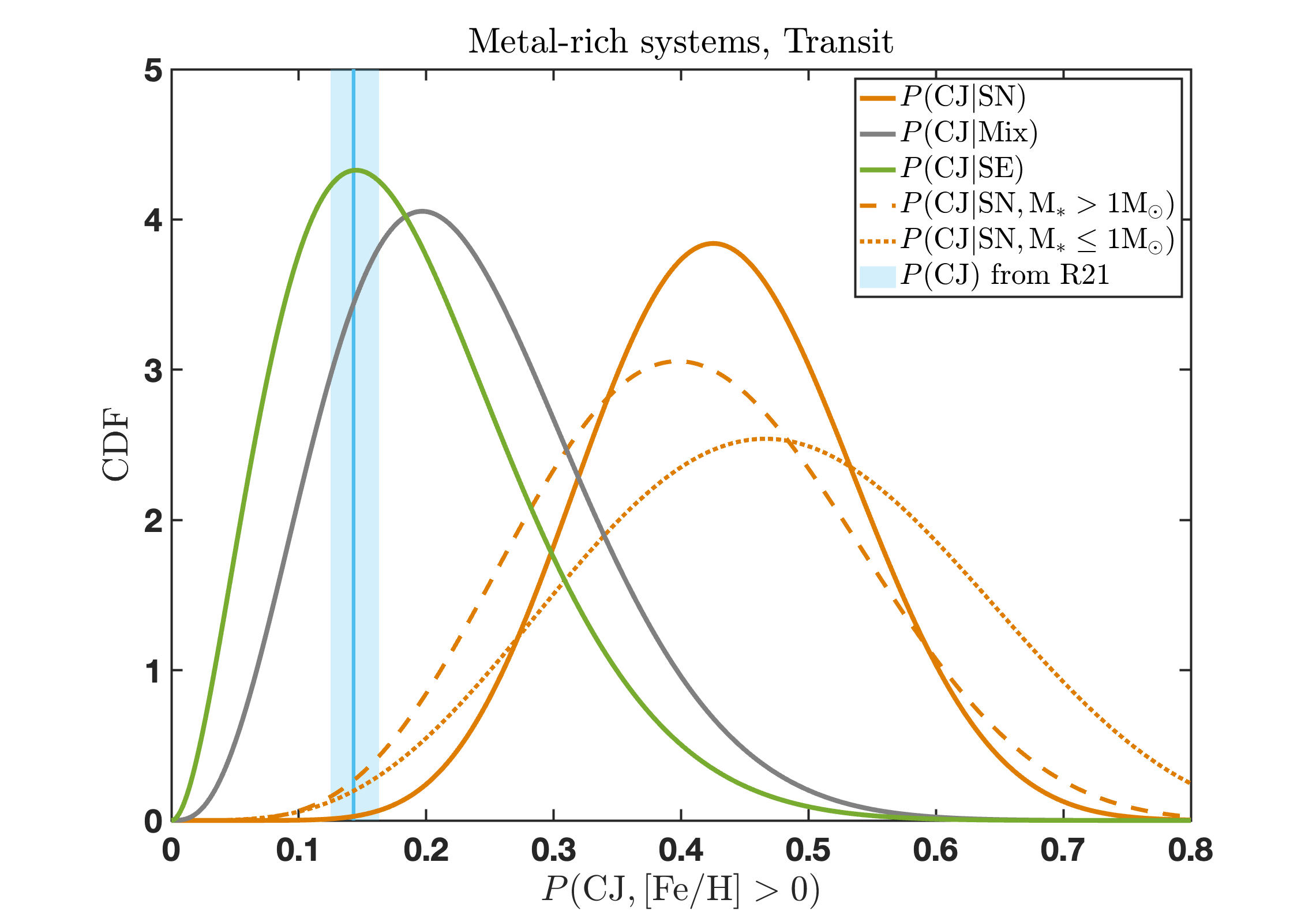}
\caption{The probability distribution functions (PDF) for the conditional frequency of cold Jupiters in transiting Sub-Neptune systems ($P (\rm CJ|SN$), orange), Mixed systems ($P (\rm CJ|Mix)$, grey) and Super-Earths systems ($P (\rm CJ|SE)$, green) around metal-rich ($\rm [Fe/H]>0$) stars.
The transiting Sub-Neptune systems were further divided into two subsamples, masses $>1 M_\odot$ (dashed orange line) and masses $<1 M_\odot$ (dotted orange line).
The solid cyan line and region denote the value and 1-$\sigma$ interval for the field giant frequency $P \ (\rm CJ)$ in metal-rich systems from R21 \citep{2021ApJS..255....8R}.
\label{figfCJ_SNMixSE_metalrich}}
\end{figure}

The transiting metal rich subsample consists of 23 sub-Neptune systems (9 systems hosting cold Jupiters), 15 super-Earth systems (2 systems hosting cold Jupiters) and 18 mixed systems (3 systems hosting cold Jupiters). 
We then calculate the posterior distributions of the conditional frequency of cold Jupiters in sub-Neptunes systems $P \rm (CJ|SN, [Fe/H]>0)$, mixed systems $P \rm (CJ|Mix, [Fe/H]>0)$, and super-Earth systems $P \rm (CJ|SE, [Fe/H]>0)$.
As shown in Figure \ref{figfCJ_SNMixSE_metalrich}, the conditional frequencies of cold Jupiter in sub-Neptunes systems is highest, followed by mixed systems, with super-Earth systems being lowest.
Specifically, $P \rm (CJ|SN, [Fe/H]>0)$, $P \rm (CJ|Mix, [Fe/H]>0)$, and $P \rm (CJ|SE, [Fe/H]>0)$ are $42.6^{+10.6}_{-9.9}\%$, $19.6^{+15.8}_{-5.2}\%$, and $14.5^{+12.7}_{-6.9}\%$, respectively.

To evaluate the correlations between inner sub-Neptunes/super-Earths and outer gas giants, we compare with the field frequency of cold Jupuiter around metal-richer stars, $P (\rm CJ, [Fe/H]>0)$.
Previous studies have provided estimates of the frequency of cold Jupiters and the correlation with stellar metallicity \citep[e.g.,][]{2010PASP..122..905J,2011arXiv1109.2497M,2021ApJS..255....8R,2022AJ....164....5Z,2023PNAS..12004179C}. 
For metal-rich stars, the frequency of cold Jupiters is $\sim 10\%-20\%$.
Here we adopt the result of \cite[hereafter R21]{2021ApJS..255....8R}, $P(\rm CJ, [Fe/H]>0) = 14.3^{+2.0}_{-1.8}\%$, as their definition of cold Jupiters in mass and semimajor axis is same to that in this work.
As can be seen in Figure \ref{figfCJ_SNMixSE_metalrich},
$P \rm (CJ|SN, [Fe/H]>0)$ and $P \rm (CJ|Mix, [Fe/H]>0)$ are larger than $P(\rm CJ, [Fe/H]>0)$, while 
$P \rm (CJ|SE, [Fe/H]>0)$ is broadly consistent with $P(\rm CJ, [Fe/H]>0)$.

To estimate the statistical significance, we assume that the conditional and field frequency follow the distributions described above, and generate 100,000 random realizations from each. 
Out of 10,000 resampled data, $P \rm (CJ|SN, [Fe/H]>0)$ is larger than $P \rm (CJ, [Fe/H]>0)$ for 9,995 times, corresponding to a positive sub-Neptune-cold Jupiter correlation in metal-rich systems with a confidence level of 99.95\%.
For mixed systems, $P \rm (CJ|Mix, [Fe/H]>0)$ is a bit higher than the field cold Jupiter frequency, however the statistical significance (84.57\%) is insufficient likely due to the small size.
Whereas, $P \rm (CJ|SE, [Fe/H]>0)$ is statistically indistinguishable from $P \rm (CJ, [Fe/H]>0)$ within $\sim 1$-$\sigma$ uncertainties, suggesting that there is no (significant) correlation between inner super-Earths and outer cold Jupiters.
The results are summarized in Table \ref{tab:FCJ}.


In addition, {\cite{2025ApJ...982L...7B} and \cite{2025A&A...700A.126B} show that the correlation between cold Jupiters and small planets appear to be restricted to high-mass stars.}
As shown in the top-left panel of Figure \ref{figRVDE_SNSEMixtransiting}, the sub-Neptune systems, mixed systems, and super-Earth systems exhibit similar stellar mass distributions. Therefore, the differences in the conditional frequency of cold Jupiters among these system types are unlikely to be driven by differences in their stellar mass distributions.
Besides, we further divided the metal-rich sub-Neptune systems into two groups according to stellar mass: $M_*>1 M_\odot$ and $M_* \le 1M_\odot$, which consists of 14 systems (5 hosting cold Jupiters) and 9 systems (4 hosting cold Jupiters).
We then calculate their conditional frequencies.
$P (\rm CJ|SN, [Fe/H]>0\&M_*>1M_\odot)$ and $P (\rm CJ|SN, [Fe/H]>0\&M_*>1M_\odot)$ are $39.7^{+13.8}_{-11.6}\%$ and $46.6^{+15.7}_{-14.8}\%$, respectively.
As shown as dashed and dotted orange lines in Figure \ref{figfCJ_SNMixSE_metalrich}, $P (\rm CJ|SN, [Fe/H]>0\&M_*>1M_\odot)$ and $P (\rm CJ|SN, [Fe/H]>0\&M_*>1M_\odot)$ show statistically consistent distributions, and both are larger than the field cold Jupiter frequency, with confidence levels of 99.07\% and 99.35\%.
Therefore, the excess of cold Jupiters in sub-Neptune systems around metal-rich stars does not appear to be restricted to high-mass stars.
For the metal-rich mixed and super-Earth systems, however, the number of systems hosting cold Jupiters is too small to allow a separate analysis.
The results are summarized in Table \ref{tab:FCJ}.

{We also examine the distribution of planet radii across metal-rich stars of different masses, and find that higher-mass stars host a larger fraction of sub-Neptune systems (14/28) and a smaller fraction of super-Earth systems (6/28), compared to lower-mass stars (for which both fractions are 9/28).
Combined with our finding of a stronger correlation between sub-Neptunes and cold Jupiters, these trends may help explain, at least in part, why metal-rich systems around higher-mass stars exhibit a higher conditional occurrence rate of cold Jupiters.
}

\begin{table*}[t]
\centering
\renewcommand\arraystretch{1.25}
\caption{Cold Jupiters frequencies.}
{\footnotesize
\label{tab:FCJ}
\begin{tabular}{l|cccc} \hline       & $N_{\rm total}$ & $N_{\rm det}$ & Conditional frequency & Field frequency \\ \hline
\multicolumn{5}{c}{Transit sample, Metal-rich ($\rm [Fe/H]>0$), Section \ref{sec.result.FCJ_rich}, Figure \ref{figfCJ_SNMixSE_metalrich}} \\ \hline
Sub-Neptune systems & 23 & 9 & $42.6^{+10.6}_{-9.9}\%$ & \multirow{5}{*}{$14.3^{+2.0}_{-1.8}\%$} \\
Sub-Neptune systems with $M_* >1 M_\oplus$ & 14 & 5 & $39.7^{+13.8}_{-11.6}\%$ &  \\
Sub-Neptune systems with $M_* \le M_\oplus$ & 9 & 4 & $46.7^{+15.7}_{-14.8}\%$ &  \\
Mixed systems & 18 & 3 & $19.6^{+15.8}_{-5.2}\%$ & \\
Super-Earth systems & 15 & 2 & $14.5^{+12.7}_{-6.9}\%$ \\ \hline
\multicolumn{5}{c}{Transit sample, Metal-poor ($\rm [Fe/H] \le 0$), Section \ref{sec.result.FCJ_poor}, Figure \ref{figfCJ_SNMixSE_metalpoor}} \\ \hline
Sub-Neptune systems & 29 & 1 & $3.9^{+6.0}_{-1.9}\%$ & \multirow{3}{*}{$5.0^{+1.6}_{-1.3}\%$} \\
Mixed systems & 11 & 0 & $0.0^{+9.7}_{-0.0}\%$ & \\
Super-Earth systems & 13 & 1 & $7.9^{+11.0}_{-3.3}\%$ \\ \hline
\multicolumn{5}{c}{Kepler-Keck sample, Metal-rich ($\rm [Fe/H] > 0$), Section \ref{sec.results.Kepler-Keck}, Figure \ref{figfCJ_SNMixSE_metalrich_KeplerKeck}} \\ \hline
Sub-Neptune systems & 7 & 3 & $45.8^{+18.6}_{-16.3}\%$ & \multirow{3}{*}{$14.3^{+2.0}_{-1.8}\%$} \\
Mixed systems & 7 & 2 & $34.8^{+19.7}_{-14.6}\%$ & \\
Super-Earth systems & 8 & 1 & $13.3^{+17.0}_{-6.8}\%$ \\ \hline
\multicolumn{5}{c}{RV sample, Metal-rich ($\rm [Fe/H] > 0$), Section \ref{sec.results.RV}, Figure \ref{figfCJ_RVSP_metalrich}} \\ \hline
Systems with $M_{\rm p}>10M_\oplus$ & 22 & 7 & $34.6^{+11.0}_{-9.1}\%$ & \multirow{2}{*}{$14.3^{+2.0}_{-1.8}\%$} \\
Systems with $M_{\rm p} \le 10M_\oplus$ & 17 & 3 & $19.1^{+11.8}_{-7.2}\%$ & \\ \hline
\end{tabular}}
\tablecomments{The field frequency is the same for all rows within each subsection.}
\end{table*}

\subsection{Conditional frequency Cold Jupiters around metal-poor stars}
\label{sec.result.FCJ_poor}

\begin{figure}[t]
\centering
\includegraphics[width=\linewidth]{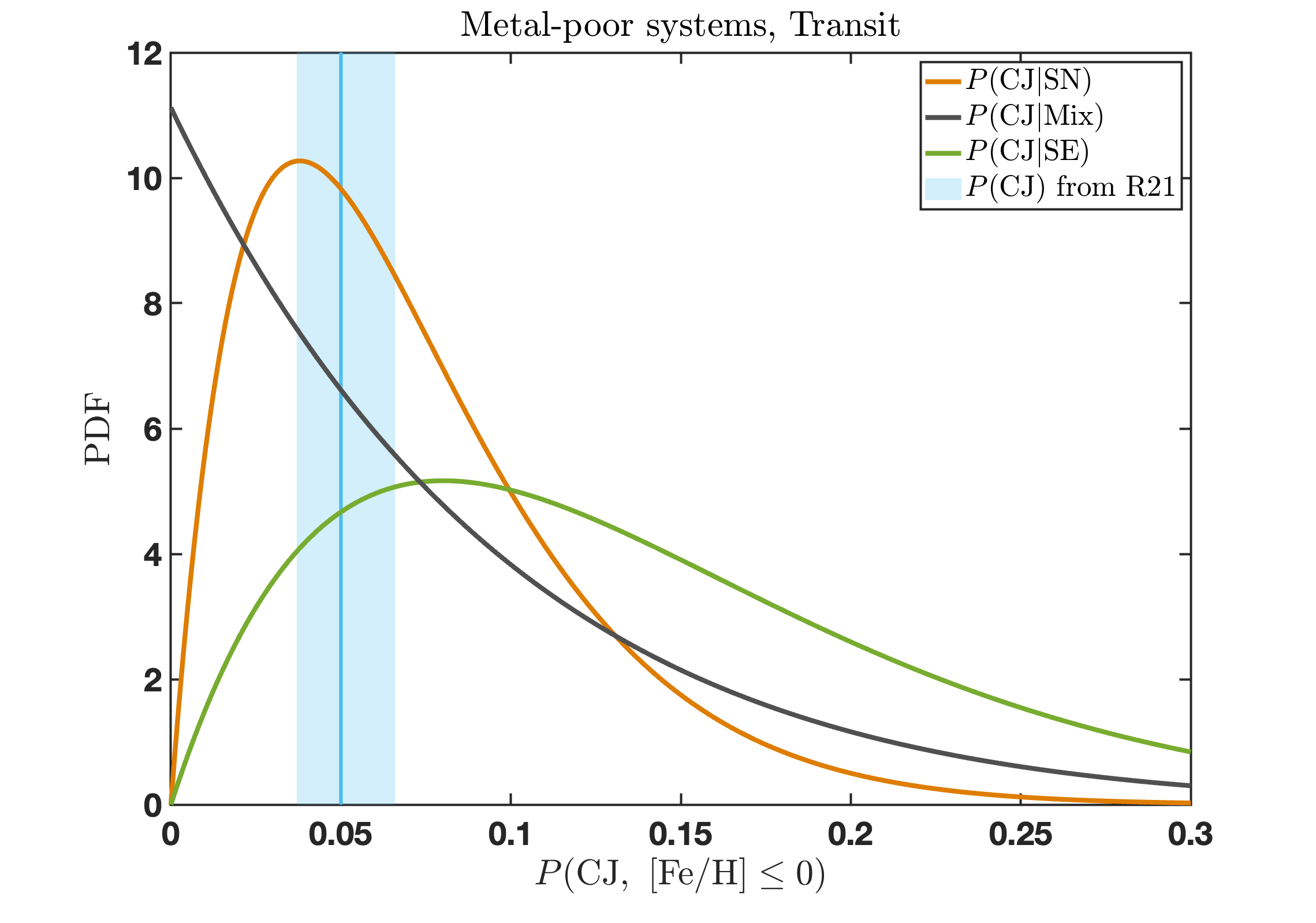}
\caption{The probability distribution functions (PDF) for the conditional frequency of cold Jupiters in transiting Sub-Neptune systems ($P (\rm CJ|SN$), orange), Mixed systems ($P (\rm CJ|Mix)$, grey) and Super-Earths systems ($P (\rm CJ|SE)$, green) around metal-poor ($\rm [Fe/H] \le 0$) stars.
The solid cyan line and region denote the value and 1-$\sigma$ interval for the field giant frequency $P \ (\rm CJ)$ in metal-poor systems from R21.
\label{figfCJ_SNMixSE_metalpoor}}
\end{figure}

For the metal-poor systems, there are 29 sub-Neptune systems (1 cold Jupiter), 11 mixed system (0 cold Jupiter), and 13 super-Earth systems (1 cold Jupiter).
Figure \ref{figfCJ_SNMixSE_metalpoor}
shows the the probability distributions their conditional frequency of cold Jupiters, which is derived with the method described in section \ref{sec.results.method}.
As summarized in Table \ref{tab:FCJ}, $P \rm (CJ|SN, [Fe/H] \le 0)$,  $P \rm (CJ|Mix, [Fe/H] \le 0)$ and 
$P \rm (CJ|Mix, [Fe/H] \le 0)$ are 
$3.9^{+6.0}_{-1.9}\%$, $0.0^{+9.7}_{-0.0}\%$, and $7.9^{+11.0}_{-3.3}\%$, respectively.
For the field frequency, we adopt the cold Jupiter frequency reported in R21, $P(\rm CJ, [Fe/H] \le 0) = 5.0^{+1.6}_{-1.3}\%$. 
$P \rm (CJ|SN, [Fe/H] \le 0)$,  $P \rm (CJ|Mix, [Fe/H] \le 0)$, $P \rm (CJ|Mix, [Fe/H] \le 0)$ are all indistinguishable with the field frequency of cold Jupiters around metal-poor star, i.e., $5.0^{+1.6}_{-1.3}\%$ within 1-$\sigma$ uncertainties, suggesting that there seem to be no significant correlation between inner small planets and outer giants.

{
\subsection{Consistent with results using Kepler-Keck subsample}
\label{sec.results.Kepler-Keck}
The transit sample analyzed in this work is compiled from multiple studies and is therefore heterogeneous, which may introduce selection biases.
For example, planets with larger radii tend to produce stronger radial velocity (RV) signals and are thus more likely to be selected for follow-up observations. Systems showing long-term RV trends may also be preferentially monitored, increasing the likelihood of detecting cold Jupiters.
However, this effect is unlikely to introduce a significant bias in our results. First, all systems in our sample include RV datasets with baselines longer than one year and more than 20 data points, which helps mitigate the selection effects described above. 
Second, the ability to detect cold Jupiters through long-term monitoring primarily depends on detection sensitivity, which is governed by the outer planet’s mass and orbital period rather than the properties of the inner transiting planet.

\begin{figure}[t]
\centering
\includegraphics[width=\linewidth]{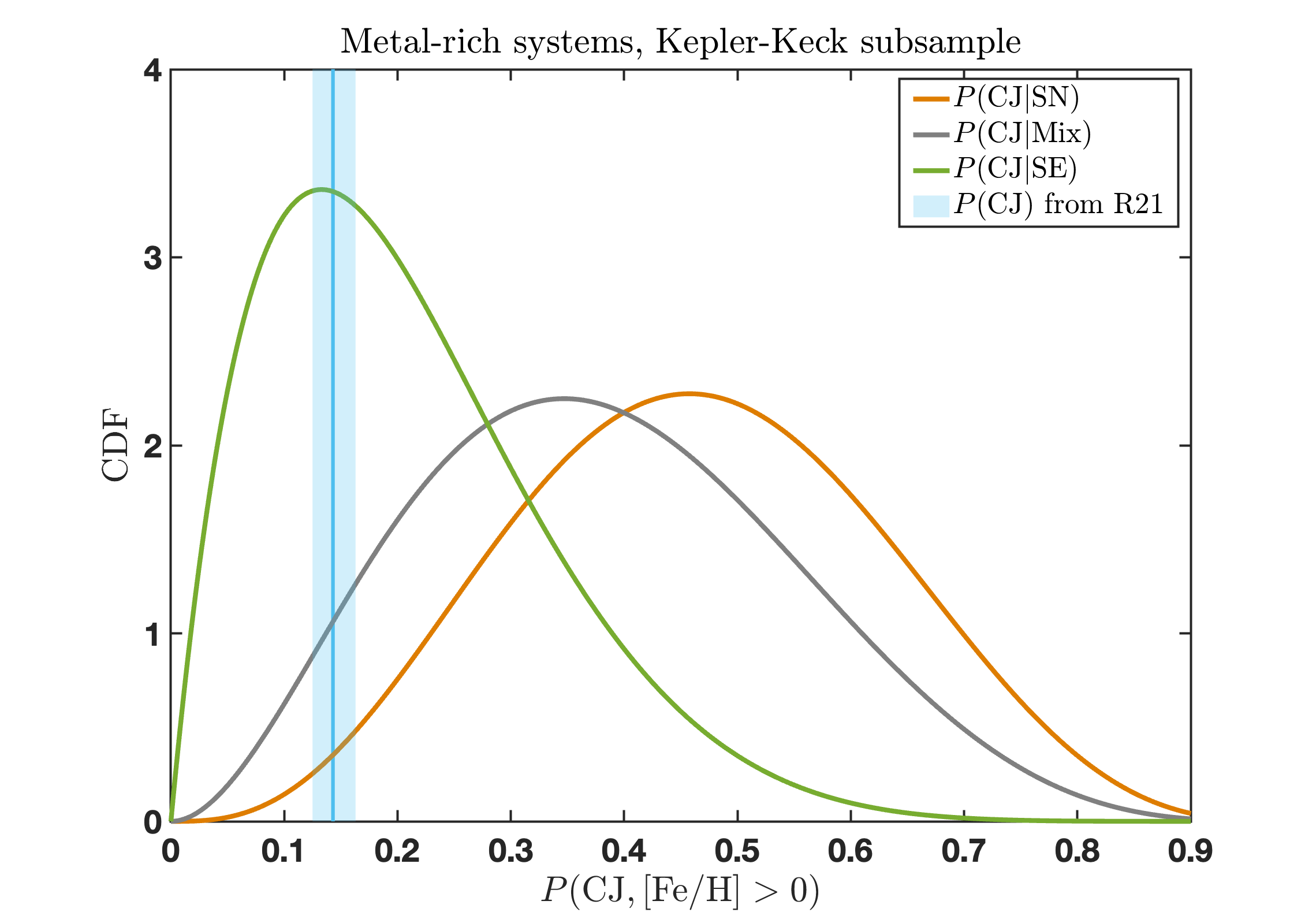}
\caption{The probability distribution functions (PDF) for the conditional frequency of cold Jupiters in Sub-Neptune systems ($P (\rm CJ|SN$), orange), Mixed systems ($P (\rm CJ|Mix)$, grey) and Super-Earths systems ($P (\rm CJ|SE)$, green) around metal-rich ($\rm [Fe/H]>0$) stars detected by the Kepler Giant Planet Survey with RV data from Keck Observatory High Resolution Echelle Spectrometer \citep{2024ApJS..270....8W}.
The solid cyan line and region denote the value and 1-$\sigma$ interval for the field giant frequency $P \ (\rm CJ)$ in metal-rich systems from R21.
\label{figfCJ_SNMixSE_metalrich_KeplerKeck}}
\end{figure}

To further assess the impact of sample heterogeneity, we construct a homogeneous subsample from the whole transit sample. 
All systems in this subsample are drawn from \cite{2024ApJS..270....8W}, with transit data obtained from Kepler and RV data acquired using the High Resolution Echelle Spectrometer (HIRES) at Keck Observatory.
This Kepler–Keck subsample consists of 22 metal-rich systems (6 hosting cold Jupiters) and 10 metal-poor systems (none hosting cold Jupiters). Among the 22 metal-rich systems, there are 7 sub-Neptune systems (3 with cold Jupiters), 7 mixed systems (2 with cold Jupiters), and 8 super-Earth systems (1 with a cold Jupiter). We then calculate their conditional occurrence rates of cold Jupiters.

As shown in Figure~\ref{figfCJ_SNMixSE_metalrich_KeplerKeck}, the conditional occurrence rate of cold Jupiters is highest for sub-Neptune systems and lowest for super-Earth systems. 
Specifically, $P \rm (CJ|SN, [Fe/H]>0)$, $P \rm (CJ|Mix, [Fe/H]>0)$, and $P \rm (CJ|SE, [Fe/H]>0)$ are $45.8^{+18.6}_{-16.3}\%$, $34.8^{+19.7}_{-14.6}\%$, and $13.3^{+17.0}_{-6.8}\%$, respectively.
Compared to the field cold Jupiter frequency, the conditional frequencies for sub-Neptune and mixed systems are higher, with confidence levels of 98.56\% and 94.01\%, respectively. 
In contrast, the conditional frequency of cold Jupiters for super-Earth systems is statistically indistinguishable from the field value.

The good consistence between the results from this homogeneous subsample and those from the full transiting sample suggests that the excess of cold Jupiters around metal-rich stars is unlikely to be driven by sample heterogeneity, but instead reflects an underlying physical trend.

}

\subsection{Consistent with results using RV sample}
\label{sec.results.RV}

\cite{2024ApJ...968L..25B} also provides 75 systems with one
or more inner small planets (i.e., $1 M_\oplus <M_{\rm P}<20 M_\oplus$) initially discovered with radial velocities (RV). 
Among them, 39 systems are hosted by metal-rich stars with 10 hosting cold Jupiters and  and 36 metal-poor systems with 1 hosting a gas giant.
To facilitate a direct comparison with the transiting sample, we divide the metal-rich RV systems into two subclasses: 
\begin{enumerate}
    \item 22 systems hosting only small planets more massive than $10 M_\oplus$, 7 of which hosting cold Jupiters.
    \item 17 systems hosting small planets with masses $
    \le 10 M_\oplus$, 3 of which hosting cold Jupiters.
\end{enumerate}

\begin{figure}[t]
\centering
\includegraphics[width=\linewidth]{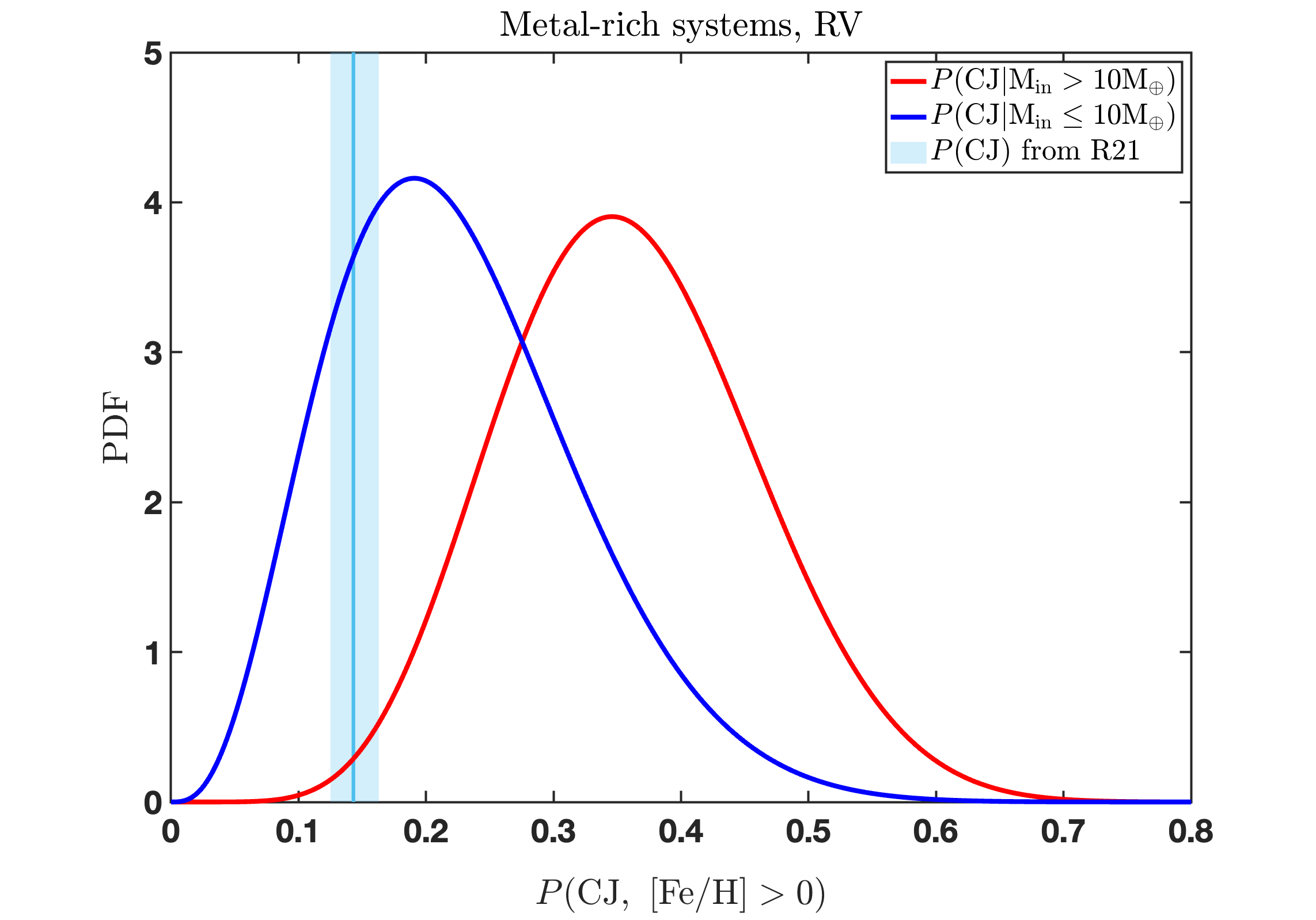}
\caption{The probability distribution functions (PDF) for the conditional frequency of cold Jupiters in RV systems hosting only massive small planets ($P \ (\rm CJ|M_{\rm p} > 10 M_\oplus$), red), and those hosting lower-mass small planets ($P \ (\rm CJ|M_{\rm p} \le 10 M_\oplus$), blue) around metal-rich ($\rm [Fe/H]>0$) stars.
\label{figfCJ_RVSP_metalrich}}
\end{figure}

Figure \ref{figfCJ_RVSP_metalrich} shows their distributions of conditional frequencies of cold Jupiters.
As can be seen, the conditional frequencies of cold Jupiters for metal-rich RV systems hosting only massive small planets $P(\rm CJ|M_{\rm in}>10 M_\oplus,[Fe/H]>0)$ is larger than that of system hosting less massive planets $P(\rm CJ|M_{\rm in} \le 10 M_\oplus,[Fe/H]>0)$.
Specifically, $P(\rm CJ|M_{\rm in}>10 M_\oplus,[Fe/H]>0)$ is $34.6^{+11.0}_{-9.1}\%$, which is larger than the field cold Jupiter frequency $P(\rm CJ, [Fe/H]>0)$ with a confidence level of 99.11\%.
In contrast, $P(\rm CJ|M_{\rm in} \le 10 M_\oplus,[Fe/H]>0)$ is $19.1^{+11.8}_{-7.2}\%$, which is a bit larger but statistically indistinguishable from $P(\rm CJ, [Fe/H]>0)$.
The above results have been summarized in Table \ref{tab:FCJ}.
Therefore, among metal-rich RV systems, a significant positive correlation with outer cold Jupiters is observed for systems containing massive inner planets, whereas such a correlation is weak or insignificant for systems hosting lower-mass inner planets.

This result is generally consistent with that obtained from the transiting sample. 
Furthermore, the RV sample is subject to stronger observational biases to smaller planets than the transit sample; 
however, these biases act to strengthen, rather than weaken, our conclusions. 
In particular, the detection of low-mass inner planets in RV systems requires high-quality data, which typically arises from intrinsically quiet host stars and/or intensive observational campaigns with a large number of measurements.
Such observing conditions in turn also enhance the detectability of outer cold Jupiters in the same systems, boosting the conditional frequency of cold Jupiters in systems hosting low-mass inner planets $P(\rm CJ|M_{\rm in} \le 10 M_\oplus,[Fe/H]>0)$.
Even with this bias, we find that $P(\rm CJ|M_p < 10 M_\oplus)$ remains low, reinforcing the conclusion that systems hosting low-mass inner planets exhibit a weak or absent correlation with outer cold Jupiters.

\section{Discussions and conclusions}
\label{sec.dis_con}
Based on 109 transiting systems with both inner small planet and outer gas giants (section \ref{sec.sample}) from \cite{2024ApJ...968L..25B}, we investigate the correlations between outer gas giants and inner planets above/below the radius valley.
We find that for metal-rich ($\rm [Fe/H]$) systems, the positive correlation is significantly strong (confidence level of 99.95\%) for systems hosting only sub-Neptunes above the radius valley, while systems hosting super-Earths show weak correlation with outer cold Jupiters (Figure \ref{figfCJ_SNMixSE_metalrich}, section \ref{sec.result.FCJ_rich}).
{We also consider a homogeneous Kepler-Keck subsample and derive a consistent trend, i.e., the strongest positive correlation between inner planets and cold Jupiters is for sub-Neptunes systems in high metallicity regime (Figure \ref{figfCJ_SNMixSE_metalrich_KeplerKeck}, section \ref{sec.results.Kepler-Keck}).} 
A similar result is observed in the RV sample, where metal-rich systems hosting massive inner planets display a strong positive correlation (confidence level of 99.11\%) with outer cold Jupiters (Figure \ref{figfCJ_RVSP_metalrich}, section \ref{sec.results.RV}).
The result suggests that stellar metallicity plays a key role in shaping planetary system architectures, influencing both the likelihood of forming outer giant planets and the mass and size distribution of inner small planets. 

Our results are consistent with several established observational trends. 
It has long been established that the frequency of cold Jupiters increases with stellar metallicity, representing the well-known planet–metallicity correlation \citep{2005ApJ...622.1102F,2010PASP..122..905J,2011arXiv1109.2497M}.
Subsequent studies have also shown that the fraction of stars hosting small planets rises gradually with metallicity \citep{2019ApJ...873....8Z}.
Moreover, the relative proportion of sub-Neptunes to super-Earths decreases as metallicity increases \citep{2018MNRAS.480.2206O,2022AJ....163..249C}. 
Recent work also shows that inner sub-Neptunes in systems hosting outer giant planets tend to exhibit systematically larger radii \citep{2025MNRAS.544L..51B}.
Our results, showing a pronounced positive correlation between massive/larger inner planets and outer cold Jupiters around metal-rich stars, extend and connect these earlier findings.

These findings can be naturally understood within the core accretion framework.
In this paradigm, protoplanetary disks around higher-metallicity stars possess more massive reservoirs of solids, which enable the rapid formation of massive planetary cores not only in the outer disk but also in the inner regions \citep{2023ApJ...952L..20C}. 
The enhanced solid content allows planetary cores to grow more efficiently and to accrete larger amounts of gas \citep{2012A&A...547A.111M,2018MNRAS.479.4786V,2021A&A...656A..69E,2025A&A...701A..94C}, thereby favoring the formation of sub-Neptunes in inner region and gas giants in outer region. 
Moreover, planets with more massive cores are better able to retain their primordial H/He envelopes against atmospheric mass-loss processes such as photoevaporation \citep{2013ApJ...776....2L,2014ApJ...792....1L} and core-powered model \citep{2018MNRAS.479.4786V}. 
As a result, inner planets in metal-rich systems are expected to preferentially populate the sub-Neptune regime above the planetary radius valley, rather than appearing as stripped super-Earths below the valley. 


{Furthermore, in metal-rich protoplanetary disks, more massive sub-Neptunes may reach the critical mass required for significant inward migration before the disk dissipates \citep{2023EPJP..138..181E,2025A&A...701A..94C}. 
Consequently, inward migration of sub-Neptunes (e.g., beyond the ice lines) could also contribute to their presence in inner planetary systems, leading to a positive correlation between inner sub-Neptunes and outer cold Jupiters \citep{2023A&A...674A.178B}.}
Metal-rich disks are capable of producing both outer cold Jupiters and inner planets with larger radii and masses, providing a coherent explanation for our statistical results in both the transiting and RV samples.

Future observations and analyses from missions such as Gaia \citep{2026AJ....171...18L}, PLATO \citep{2014ExA....38..249R}, and ET \citep{2024ChJSS..44..400G} will further test our results and help provide a more definitive assessment of the impact of cold Jupiters on inner small planets.

\begin{acknowledgments}
We thank Wei Zhu for helpful discussions and suggestions.
This work is supported by the National Key R\&D Program 
of China (Grant No. 2024YFA1611803), the National Natural Science 
Foundation of China (Grant No.12403071, 12373071) and the science research grants from the China Manned Space Project (CMS-CSST-2025-A16, CMS-CSST-2021-B09, CMS-CSST-2021-B12, and CMS-CSST-2021-A10).
\end{acknowledgments}



\appendix
\renewcommand\thefigure{\Alph{section}\arabic{figure}}
\setcounter{figure}{0} 
\renewcommand\thetable{\Alph{section}\arabic{table}}
\setcounter{table}{0} 

\section{Detection completeness for the three different types of transiting systems}

\begin{figure*}[h]
\centering
\includegraphics[width=\linewidth]{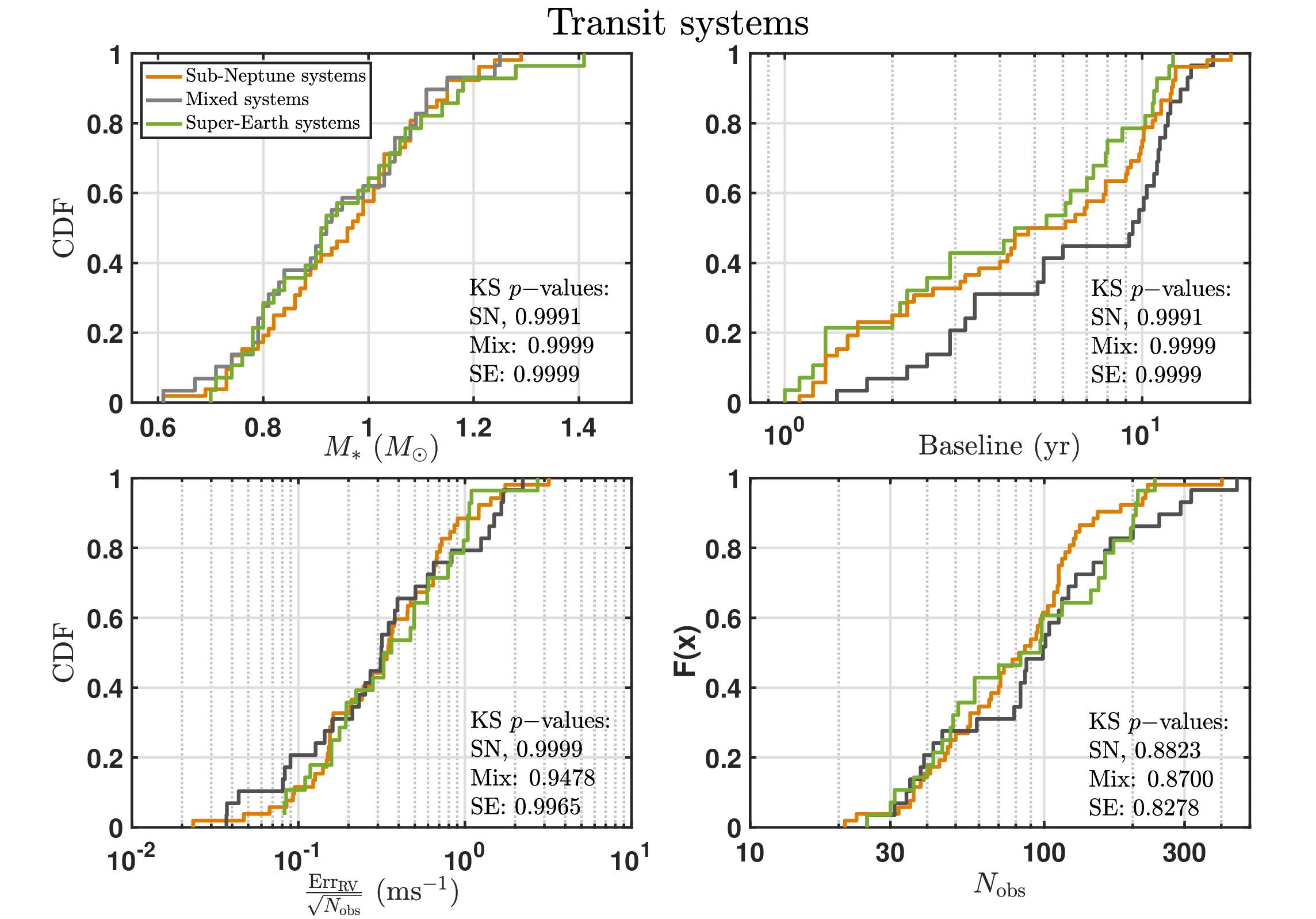}
\caption{The cumulative distributions of stellar mass (Top-Left),  observational times baselines (Top-right), the uncertainties in radial velocity (RV) measurements divided by $\sqrt{N_{\rm obs}}$ (Bottom-Left), and the number of RV data points ${N_{\rm obs}}$ (Bottom-right) for the three types of planetary systems.
In the bottom-right corner of each panel, we print the KS $p-$values between the three types of systems and the whole transiting sample.
\label{figRVDE_SNSEMixtransiting}}
\end{figure*}

As described in Appendix A of \citep{2024ApJS..270....8W}, the detection efficiency of outer cold Jupiters depends primarily on the RV measurement uncertainties and the temporal baseline of the observations (summarized in Table \ref{tab:Transitsystem_catalog}). Therefore, we collected the RV measurements, their associated uncertainties, and the observational baselines from the original literature. {Figure \ref{figRVDE_SNSEMixtransiting} shows the distributions of stellar mass, observational times baselines, the uncertainties in radial velocity (RV) measurements divided by $\sqrt{N_{\rm obs}}$, and the number of RV data points ${N_{\rm obs}}$ for three different types of planetary systems (i.e., sub-Neptune, mixed, and super-Earth) within the transit sample.}
We further applied Kolmogorov–Smirnov (KS) tests to compare the distributions of each system type with that of the full transit sample. 
In all cases, the resulting KS test p-values are larger than 0.3, indicating that the distributions are statistically consistent. Consequently, the detection completeness among the three types of systems does not differ significantly from a statistical perspective.

\begin{deluxetable*}{lccc}[!h]
\tablecaption{Catalog of transiting systems} \label{tab:Transitsystem_catalog}
\tabletypesize{\scriptsize}
\tablewidth{0pt}
\tablehead{
\colhead{Target} & \colhead{$N_{\rm obs}$} & \colhead{Baseline} & \colhead{RV uncertainty}  \\
 & & \colhead{[year]} & \colhead{$\rm  m s^{-1}$}  \\
} 
\startdata
CoRoT-24 & 71 & 3.2 & 13.9 \\
EPIC 220674823 & 101 & 3.4 & 4.0 \\
EPIC 229004835 & 126 & 2.3 & 3.5 \\
EPIC 249893012 & 99 & 1.4 & 2.3 \\
GJ 143 & 112 & 15.2 & 1.3 \\
GJ 9827 & 128 & 10.8 & 1.6 \\
HD 137496 & 172 & 2.5 & 1.4 \\
HD 15337 & 87 & 13.7 & 1.2 \\
HD 183579 & 56 & 7.9 & 1.2 \\
HD 18599 & 103 & 4.8 & 1.4 \\
HD 191939 & 182 & 1.3 & 1.7 \\
HD 207897 & 122 & 17.7 & 1.7 \\
HD 213885 & 238 & 10.2 & 3.0 \\
HD 23472 & 104 & 1.7 & 0.4 \\
HD 307842 & 58 & 2.1 & 4.5 \\
HD 3167 & 452 & 5.3 & 0.9 \\
HD 39091 & 402 & 2.2 & 0.5 \\
HD 5278 & 41 & 1.1 & 0.4 \\
HD 73583 & 118 & 3.5 & 1.7 \\
HD 80653 & 208 & 2.9 & 1.7 \\
\enddata
\tablecomments{
A subset of the systems listed in the table is shown here. The full catalog is available in machine-readable form online.}
\end{deluxetable*}


\clearpage
\bibliography{sample701}{}
\bibliographystyle{aasjournalv7}

\end{document}